# Rapid Sensing of Heat Stress using Machine Learning of Micrographs of Red Blood Cells Dispersed in Liquid Crystals

Prateek Verma[1], Elizabeth Adeogun[1], Elizabeth S. Greene[2], Sami Dridi[2], Ukash Nakarmi[3], Karthik Nayani[1, *]

[1]Department of Chemical Engineering, University of Arkansas, Fayetteville, Arkansas, USA

[2]Department of Poultry Science, University of Arkansas, Fayetteville, Arkansas, USA

[3]Department of Computer Science and Computer Engineering, University of Arkansas, Fayetteville, Arkansas, USA

*KEYWORDS. Health sensors, convolutional neural networks, machine learning, red blood cells, liquid crystals, soft matter, poultry.*

**ABSTRACT:** An imbalance between bodily heat production and heat dissipation leads to heat stress in organisms. In addition to diminished animal well-being, heat stress is detrimental to the poultry industry as poultry entails fast growth and high yield, resulting in greater metabolic activity and higher body heat production. When stressed, cells overexpress heat shock proteins (such as HSP70, a well-established intracellular stress indicator) and may undergo changes in their mechanical properties. Liquid crystals (LCs, fluids with orientational order) have been recently employed to rapidly characterize changes in mechanical properties of cells enabling a means of optically reporting the presence of disease in organisms. In this work, we explore the difference in the expression of HSP70 to a change in the LC response pattern via the use of convolutional neural networks (CNNs). The machine-learning (ML) models were trained on hundreds of such LC-response micrographs of chicken red blood cells with and without heat stress. Trained models exhibited remarkable accuracy of up to 99% on detecting the presence of heat stress in unseen microscope samples. We also show that crosslinking the chicken and human RBCs using glutaraldehyde in order to simulate a diseased cell was an efficient strategy for planning, building, training, and evaluating ML models. Overall, our efforts build towards the rapid detection of disease in organisms, which is accompanied by a distinct change in the mechanical properties of cells. We aim to eventuate CNN-enabled LC-sensors can rapidly report the presence of disease in scenarios where human judgment could be prohibitively difficult or slow.

Heat stress (HS) occurs when an animal is unable to regulate its body temperature in response to high environmental temperature, resulting in hyperthermia (increased body temperature). HS is detrimental to the well-being of an animal, causing discomfort, organ damage, or even death. In livestock and poultry industry, HS is known to lead to massive economic losses in addition to decreased welfare of the animals [1]. Increasing global temperatures due to climate change and ever-increasing demand for meat production have prompted research efforts toward better understanding the effects of heat stress and ways to alleviate them [2, 3]. HS is a particularly important stressor for the poultry industry, as poultry entails fast growth and high yield, resulting in greater metabolic activity, higher body heat production, and decreased thermo-tolerance [1, 4-7]. In fact, it is estimated that the amount of metabolic heat produced by the modern broiler has increased by 30% over the last 20 years [3]. In poultry, study of HS and its effect on feed intake [4, 8, 9], immunosuppression [10-12], growth [4, 7], gut health [4, 5, 9], and meat yield [4, 7, 13], etc., as well as its effect on physiological responses, such as increased production of heat shock proteins (HSPs, such as HSP70) [14, 15] or any other biomarker (such as GRP75 or Orexin) [16-18], have gained momentum recently. Facile methods to rapidly characterize the health of poultry and livestock are important in a broad range of contexts, which include understanding their health/stress status, welfare, and prediction of diseases and stressors [1, 3]. However, there remains a wide knowledge gap in coupling molecular/protein signatures of disease/stress to rapid readouts. Aside from economic concerns, overall animals' well-being is greatly diminished by HS and has become a prominent concern for consumers. Therefore, there is an urgent need to develop rapid-reporting methods that can inform on whether the organism is experiencing HS. Here, we introduce the concept of rapidly characterizing the mechanical properties of red blood cells (RBCs) of chickens using fluids called liquid crystals (LCs, fluids with orientational order). A key aspect of the development of our LC-based platform involves building ML-based convolutional neural networks (CNNs) that can generate classifiers to separate image sets of RBCs dispersed within LCs into healthy ones and those of chickens experiencing HS [19-21].

The fundamental hypothesis that drives our research is that the cells overexpressing well-established intracellular stress chaperones such as heat shock proteins (HSP70) also undergo cellular changes, for instance, the mechanical properties of the cellular membrane, which, in turn, can be detected by dispersing them in LCs. The expression levels of HSP70 will be used to define our classes for the CNN framework we develop. Current methods for monitoring stress rely on the identification of molecular and protein markers such as corticosterone and HSPs [22, 23]. Although methods that report on molecular and protein markers have increased our understanding of HS, these methods are usually time-intensive and are not immediately accessible to the end user (farmer, technician on a production line) seeking to make informed decisions on the health and stress levels of chicken. Therefore, there is a critical need to identify reliable and rapid ways to monitor HS in poultry[22, 23].

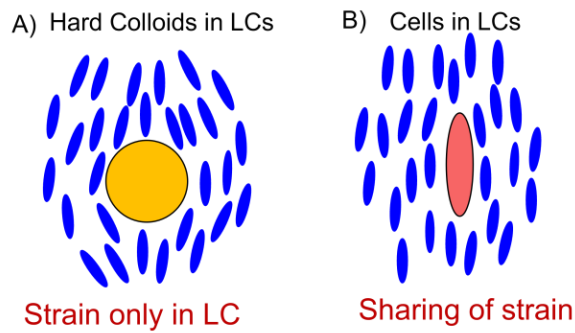

Figure 1: Schematic depicting the fundamental principle involved in our work. A) A hard colloid such a silica particle induces strain in the LC fluid B) A soft cell membrane stretches along the LC direction and releases the strain

A key innovation in this work and our methodology is to connect the expression of HSP70 to rapid optical readouts, which characterize the health of the blood cells of chickens. Previously, an LC-based technique has been deployed to rapidly report on the health of human RBCs[24]. The underlying principle is depicted in Figure 1. Molecules of LC (blue ellipsoids) are perturbed from the preferred parallel orientation when an inclusion, for instance, a colloidal particle, is present within the LC fluid (blue ellipsoids bend around the yellow particle in Figure 1A). This creates an orientational strain within the LC, as depicted in Figure 1A. However, if the inclusion is soft, such as an RBC, the LC can stretch out the cell and release some of the strain contained within the fluid. This sharing of strain is intimately coupled with the mechanical properties of the RBCs, which we expect to change as they experience HS. LCs enable rapid readouts of the mechanical properties of cells, for instance, a simple experiment of dispersing a few µl of blood in LCs can be used to understand the health status of over a thousand cells within a few minutes[24, 25].

Physiological mechanisms of chickens' response to HS or to any 'cure' employed to fight HS are far from understood. Such studies require controlled and careful broiler studies, spanning weeks, and more often than not, the blood (for elaborate examination of genes or the biomarkers) or even sacrificing the chickens. Steps are being taken towards non-invasive examination of HS, such as using feather HSP70 (a specific HS protein) [15]. To help with this, the authors ideate that the wealth of existing information, and more easily obtainable information, could be put to good use by training ML algorithms to aid in rapid identification of HS, HS biomarkers, HS susceptibility of various chicken subspecies, effectiveness of HS treatments, and so on.

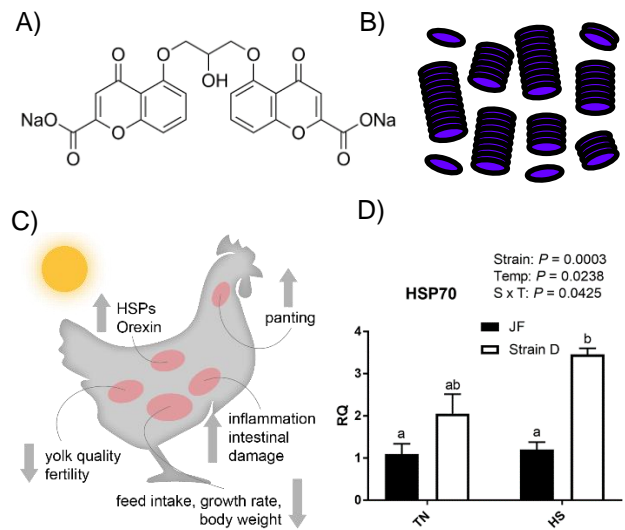

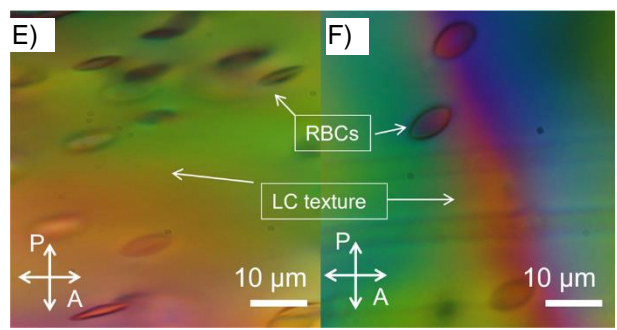

Figure 2. (A) a disodium cromoglycate (DSCG) liquid crystal molecule. (B) DSCG molecules, represented by a purple oval, stack to form a liquid crystal phase. C) Qualitative effects of heat stress on chicken and D) relative expression of HSP70 determined using the 2-ΔΔCT method, with normalization to 18s expression. (E and F) Optical Micrograph of red blood cells (RBCs) dispersed in disodium cromoglycate (DSCG) liquid crystal from E) Cobb 700 strain and F) jungle fowl. Imaging was performed in cross-polarized mode.

The motivation for this work lies in our initial observation that there was a dramatic difference in the extent of the strain of RBCs of modern-day broiler chickens and their jungle fowl ancestors, as presented in Figure 2. The LC we use is disodium cromoglycate, whose disc-shaped molecular structure is shown in Figure 2A[24, 26, 27]. The self-assembly of the disc-like DSCG molecules into rod-like stacks is depicted in Figure 2B [24, 26, 28-30]. Figure 2E and 2F shows the dramatic difference in the straining of RBCs of commercial boiler strain chicken (Cobb 700) versus that of south-east jungle fowl in DSCG.

In this study, to simulate an unhealthy RBC, whose mechanical properties differ from those of a healthy RBC, glutaraldehyde was used to crosslink the RBCs and stiffen them. Consequently, the magnitude of stretch in a glutaraldehyde treated RBC in aqueous DSCG would be lower or even absent. Optical micrographs of healthy RBCs in DSCG and of crosslinked RBCs in DSCG were used to train a simple convolutional neural network (CNN)[31]. Eventually, chicken RBCs expressing HSP70 were dispersed in LCs to test our hypothesis and confirm whether HS could be detected through visual observation as well as through CNN classification.

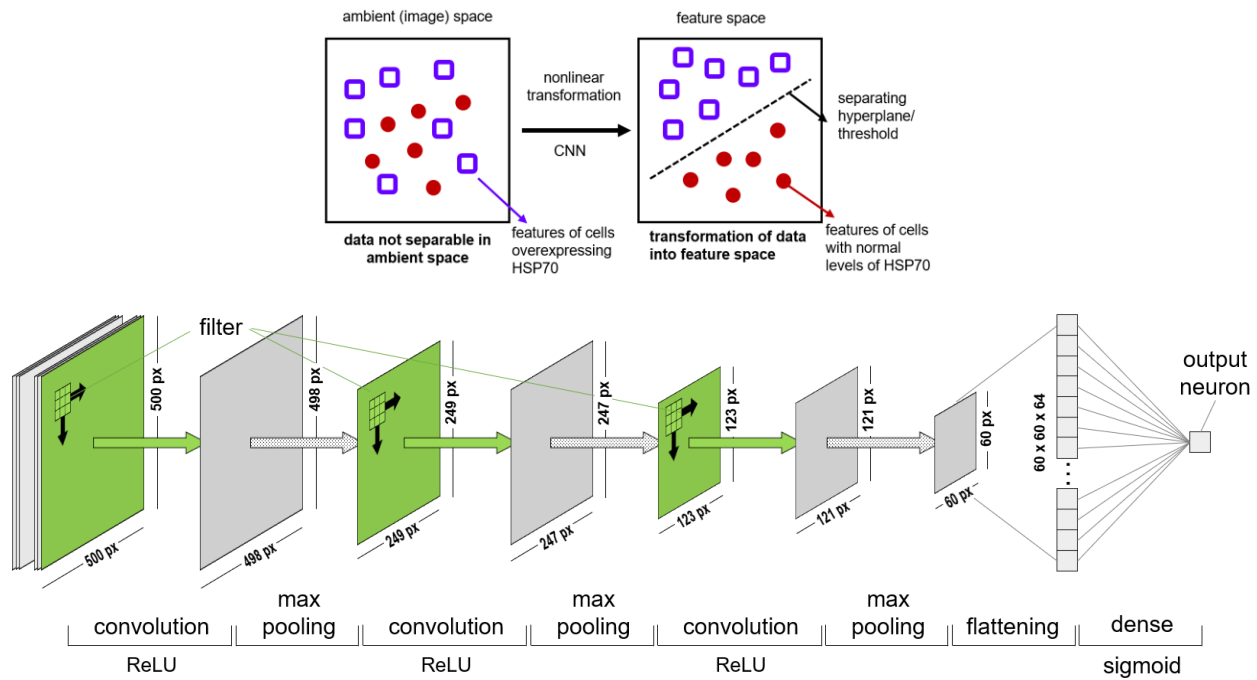

Figure 3. (A, top) A hyperplane obtained through nonlinear transformation during neural network training helps separate data into different classes. Inseparable data (squares and circles) are made separable after the calculation of the hyperplane. (B, bottom) The CNN-based machine learning model used in this study. Data in image pixels flows from left to right through three convolutional (green) and three pooling (gray) layers and culminating in a single output value. A 3x3 kernel, pixel sizes of images, and activation functions used in this study are noted in the figure.

CNNs have emerged as ideal machine learning architecture for classification of images. Image classification through CNNs work by identifying and separating critical information (features) in the image using nonlinear convolutional operation and finding an optimum hyper plane/threshold for classification (Figure 3A). Since the advent of architectures such as AlexNet [32] and VGGNet [33], classification of images has become faster and more accurate. CNNs consist of convolutional layers, that perform a mathematical convolution operation on the incoming image using a small filter (also called 'kernel') of size such as 3 x 3 pixels (shown in green in Figure 3B). CNNs learn by optimizing the values of the filter which results in correct identification of the images. Sets of images with known labels (also called 'classes') are fed through the CNNs repeatedly for learning until, a good enough accuracy is achieved. Through convolutional learning, CNNs have shown to be able to detect edges, shapes, and other, sometimes imperceptible, features of an image that enables them to perform the classification. Figure 3B shows progression of data through a typical CNN, composed of convolutional layers, each followed by a max-pooling layer (that reduces the 2D-image size) ending in a single output (for binary classifications) that denotes the probability of the data belonging to one out of the two classes.

## EXPERIMENTAL SECTION

### Materials

Human RBCs (extracted from whole blood) were purchased from Innovative Research Inc. (Novi, Michigan, USA). Chicken RBCs were extracted in accordance with approved IACUC Protocol #21050. A 154 mM solution of NaCl was prepared for dispersing RBCs. Cross-linking of RBCs was performed using glutaraldehyde (Grade I, 25% aqueous) purchased from Sigma Aldrich. Disodium cromoglycate (DSCG) was purchased from TCI America (Portland, Oregon, USA). The molecular structure of DSCG and its LC stacking have been shown in Figure 3. Deionized water with a resistivity of 18.2 MΩ.cm was obtained using a Milli-Q system (by Millipore) and was used wherever water was required.

### Computational:

All computations and programs were run on a Linux machine running Ubuntu 20.04 LTS using hardware consisting of an i7-11700K 3.6 GHz CPU, 32 GB of DDR4 3200 MHz RAM, and a GTX 1660 Ti GPU. Python (version 3.9.5) was primarily used for programming. Within Python, TensorFlow library using Keras API was used to build and train neural network models; image processing was performed using the Python Image Library (PIL); and numerical data were primarily plotted using Matplotlib library.

Sample preparation and optical microscopy:

As-received RBCs were dispersed in a 154 mM NaCl isotonic solution; an isotonic solution ensured that cells maintained their natural elliptical shape. Typically, about 5 μL RBCs were mixed in about 60 μL isotonic NaCl solution to obtain the RBC dispersion. A 17.3% (w/v) DSCG aqueous solution was prepared by mixing DSCG in water in a vortex mixer for 4 hours. Previous works have shown that 17.3% (w/v) aqueous DSCG is isotonic with the interiors of an RBC, which ensured that the RBC shape change was solely due to the mechanical interaction between the LC field and the RBC.

Since RBCs naturally strain when put in a DSCG solution, strained cell samples were obtained by adding 2 μL of dispersed RBCs to 60 μL of DSCG solution and gently swirled. To prevent straining in DSCG and obtain crosslinked cell samples, glutaraldehyde was used to crosslink and stiffen the RBCs. A stock solution of 5% v/v of glutaraldehyde in water was used; 5 μL RBCs were slowly pipetted into 0.2 μL of this stock to effect crosslinking. The final glutaraldehyde concentration in the cells was chosen to be around 0.2 % to make sure the individual cells were fixated and do not form aggregates. The solution was slowly mixed on a shaker for an hour to allow the glutaraldehyde to completely crosslink. About 2 μL of crosslinked RBC was then added to 60 μL of DSCG solution and gently swirled. HSP70 RBCs were collected from 21-day old broiler chickens that had been exposed to acute heat stress (35°C for 2 hours). Whole blood was collected into EDTA coated tubes and the RBCs were isolated from the whole blood by centrifugation and washing with PBS three times.

For imaging, RBC samples were transferred (post swirling) to microscope slides. Micrographs were obtained using an Olympus BX41 optical microscope fitted with a 40x objective lens. Polarized and brightfield micrograph images were captured in the presence and absence of a polarizer respectively.

Building datasets

Images from the microscope(s) were obtained in a variety of sizes, aspect ratios and formats. Images with differences in tint, brightness, contrast, lighting were included. Images with scratch marks on the microscope slides or of samples containing foreign objects (like dirt or lint) or containing things other than the RBCs were also included. This was done to increase the diversity in the dataset, keeping in mind the plausible diversity that the trained model may encounter during testing and after deployment.

Collected images were at least 1800 px in height, either 3:2 or 4:3 in aspect ratio (width to heigh ratio), and saved in one of jpeg, bmp, tiff, or raw format. All images were RGB (containing information in red, green, and blue channels). Before building the dataset, all images were cropped and resized to the same size and converted to jpeg format. A square section from the center of the image was selected. This was done to (1) discard the sides which sometimes contained portion of the microscope slide or stage outside of the actual sample and (2) standardize the aspect ratio to 1:1 from 3:2 or 4:3. The resulting square image was scaled down in size to exactly 1000 px wide and 1000 px tall using bicubic resizing algorithm in PIL.

Each image was assigned labels (such as species, chemicals used, and magnification etc.) that were stored in a tabular form within excel files. These labels were used to programmatically find images matching a certain criterion. For instance, polarized images of crosslinked chicken RBCs were found by logical querying of these labels: polarized is "True", chemicals used contains "glutaraldehyde", species is "chicken", and cell type is "RBC".

**Table 1. Split of images between training, validation and test set in each dataset**

| class or dataset name | train | val | test |
|---|---|---|---|
| brightfield micrograph | | | |
| A. crosslinked chicken | 112 | 24 | 24 |
| B. healthy chicken | 585 | 125 | 125 |
| C. crosslinked human | 350 | 75 | 75 |
| D. healthy human | 350 | 75 | 75 |
| E. HS chicken | 489 | 105 | 105 |
| polarized | | | |
| F. crosslinked chicken | 798 | 171 | 171 |
| G. healthy chicken | 1264 | 271 | 271 |
| H. crosslinked human | 390 | 83 | 83 |
| I. healthy human | 350 | 75 | 75 |
| J. HS chicken | 423 | 90 | 90 |

Ten distinct subsets of the dataset (called classes) were built using this process and are summarized in Table 1. Each class was shuffled and split into training (`train`), validation (`val`), and test (`test`) sets in 70:15:15 ratio, respectively.

**Building the CNN:**

Convolutional neural networks (CNNs) were built using the TensorFlow library (that utilizes Keras API) in Python. Images from the datasets were imported using flow_from_directory function of Keras's preprocessing module, which yields batches of images indefinitely during training. The preprocessing module was also set to randomly flip the images horizontally or vertically during training, to add to the richness of the dataset as part of the data augmentation process. The batch size was set to 32 and the imported image size was set to 500 x 500 pixels.

**Table 2. Summary of the CNN architecture obtained using Keras's summary method after defining the model. The type, shape of the output tensor, and the number of parameters (params) for each Keras layer are shown.**

| layer name (type) | output shape | params |
|---|---|---|
| conv2d_1 (Conv2D) | (None, 498, 498, 16) | 448 |
| maxPooling-1 (MaxPooling2D) | (None, 249, 249, 16) | 0 |
| conv2d_2 (Conv2D) | (None, 247, 247, 32) | 4640 |
| maxPooling-2 (MaxPooling2D) | (None, 123, 123, 32) | 0 |
| conv2d-3 (Conv2D) | (None, 121, 121, 64) | 18496 |
| maxPooling-3 (MaxPooling2D) | (None, 60, 60, 64) | 0 |
| flatten_1 (Flatten) | (None, 230400) | 0 |
| dense_1 (Dense) | (None, 1) | 230401 |

Total params: 253,985

Trainable params: 253,985

A simple binary classifier CNN, meaning it classified images into one of two input classes, was built. The model is schematically shown in Figure 2b. It consisted of three convolutional layers, each followed by a max-pooling layer. A ReLU activation function was used with each convolutional layer. Image tensors were flattened to a single dimension before being passed to the final dense layer consisting of a single neuron; this neuron yielded the probability that a particular image belonged to one of the two classes. This probability was yielded by a sigmoid activation function, that maps the input to a value between 0 and 1. The model's parameters are shown in Table 2. The model was compiled using the optimizer Adam and the loss function sparse_categorical_crossentropy. The performance of the model was measured using accuracy as the metric.

**Training the CNN:**

In a given machine-learning experiment, a model could be trained to distinguish between two of any of the classes shown in Table 1. Specifically, models were trained to distinguish between crosslinked and healthy cells ('A and B', 'C and D', 'F and G', 'H and I') and between healthy and HS chicken cells ('B and E', 'G and J'). Experiments are depicted as first initials of class names (such as 'AB', 'CD' and so on) used in it, for convenience. Training was performed using the `fit` function on training image set, and accuracy was calculated and recorded during training for both training and validation sets. A receiver operating characteristic (ROC) curve was plotted for each experiment for a range of decision thresholds (101 linearly spaced threshold values from 0 to 1) for predictions made on validation sets. The best threshold was selected to be the one that was closest to the point (0, 1) on the ROC plot. The confusion matrix corresponding to the best threshold has been reported. A confusion matrix was also calculated for the predictions made on the test set using the best threshold (that was evaluated on the validation set) and has been reported.

Fine-tuning hyperparameters

Hyperparameters – batch size and learning rate (for `Adam`) were found to be optimal (fastest conversion to maximum validation accuracy) when set to 32 and 0.0001, respectively. This optimization was performed for the experiment 'AB' and 'GH' and the hyperparameters were set constants for all other experiments as they resulted in adequate convergence rates.

## RESULTS AND DISCUSSION

**Datasets:**

One representative microscope image from each class has been shown in Figure 4. Brightfield micrograph images (classes A-E) and corresponding polarized images (classes F-G) are shown side-by-side. Figure 4A through Figure 4J shows optical micrographs of cells suspended in nematic LC phases of 17.3 wt% DSCG at 25 °C. Inspection of Figure 4 reveals that the RBCs in the nematic phase of DSCG assume extended shapes with major axes aligned parallel to the far-field orientation of the LC (white arrow in Figure 4). Cells crosslinked with glutaraldehyde (A, C, F, H) before suspending in the nematic phase of DSCG were observed to be less strained than corresponding cells that were not crosslinked (B, D, G, I). In the selected images, this was evident when looking at the lengths, widths, and the shapes of the cells; the length and the aspect ratio of the strained (and therefore, healthy) cells were larger than those of the crosslinked cells. Between chicken and human RBCs, it was observed that the strained (i.e., healthy) chicken cells (B, G) had a higher aspect ratio than that of human cells (D, I). This was also true for crosslinked/crosslinked chicken (A, F) and human cells (C, H). Notably, the shape difference between crosslinked and healthy chicken cells (A vs B, F vs G) was not as pronounced as the difference between crosslinked and healthy human cells (C vs D, H vs I). Lastly, heat-stressed chicken RBCs were observed to have a strain somewhere between the healthy (i.e., crosslinked) (B, G) and crosslinked (A, F) chicken RBCs. This observation was consistent with our hypothesis that heat stress causes changes in a healthy cell that can manifest as mechanical stiffening.

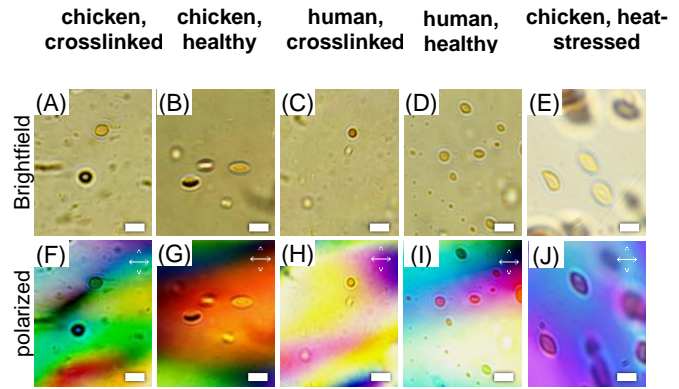

Figure 4. Brightfield images of (A) crosslinked chicken red blood cells, (B) healthy chicken red blood cells, (C) crosslinked human red blood cells (D) healthy human red blood cells (E) heat stressed chicken red blood cells. Cross-polarized images of: (F) crosslinked chicken red blood cells, (G) healthy chicken red blood cells, (H) crosslinked human red blood cells (I) healthy human red blood cells (J) Heat stressed chicken red blood cells. The crosslinked cells were obtained by crosslinking with glutaraldehyde. Scale bar represents 20 μm.

Classification of healthy and crosslinked RBCs:

Figure 5 shows the results from experiments AB, CD, FG, and HI that aimed towards the classification of healthy RBCs and crosslinked (unhealthy) RBCs submerged in LC for both chickens and humans. These experiments served as a prelude to the core objective of this work, i.e., the successful identification of cells experiencing HS. These experiments served at least three important purposes – (1) establish the ability of CNNs to successfully identify straining of cells in LCs, (2) establish generalizability of this approach towards identification of cells with different mechanical properties not only due to HS, but due to any biological response, and (3) assess the applicability of this approach to species other than chicken. Top to bottom in Figure 5, the panels show the accuracies and the loss calculated on the training and the validation sets across 100 epochs of training, followed by the ROC curve and the confusion matrix obtained for the validation set, followed, finally, by the confusion matrix obtained for the test set. A validation accuracy of 100% was obtained for experiments using unpolarized images and greater than 97% for the experiments using polarized images at the end of the training. While the loss calculated for the training and validation sets were close to each other and reached zero for the unpolarized experiments, the loss for the validation set did not drop in sync with the training set for the polarized experiments and did not reach zero.

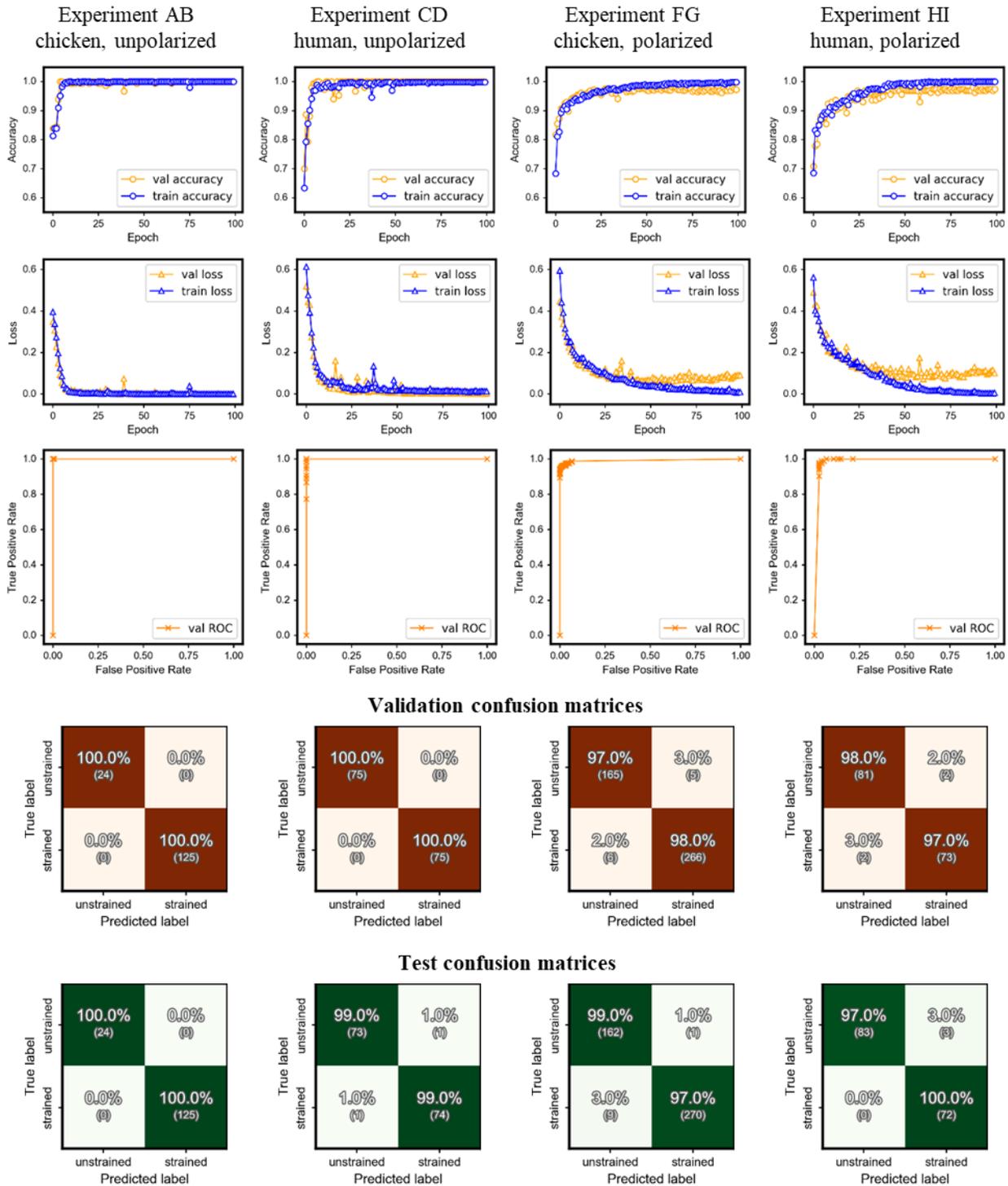

Figure 5. Results from experiments for the classification of strained (healthy) and crosslinked (unhealthy) chicken or human RBCs carried out on datasets A and B (column AB), C and D (column CD), E and G (column EG), and H and I (column HI). Each column (top to bottom) shows accuracy achieved on training and validation sets over 100 epochs of training; loss on training and validation sets over 100 epochs of training; ROC curve for validation set; confusion matrix for the validation set for the best threshold established from the ROC curve; confusion matrix for the test set obtained using the best threshold obtained on the validation set.

Clearly, polarized images posed a greater challenge to CNN. Polarized images have richer information concerning the cellular response in an LC field. The color gradients around a cell are representative of the LC director being disrupted by the presence, the shape, and the physical properties of the cell. These color gradients, that are absent in unpolarized images, may contain information about the cell's physical properties, its biochemical signatures, and hence its health, that is not readily perceptible to human visual analysis. Longer range color gradients visible in the polarized images, that are not centered on the cell and not perturbed by the presence of the cells or foreign objects (like dust or a piece of fiber), likely don't contain information linked directly to the cell's response. Because the data labels for crosslinked or healthy samples were derived solely from the fact whether glutaraldehyde was used or not, CNN's accuracy denoted the level at which it could successfully map the complex information in the images (especially the polarized ones) to the effects of glutaraldehyde on the cells. In this light, the accuracy of 97% was remarkable, given the complexities in the images.

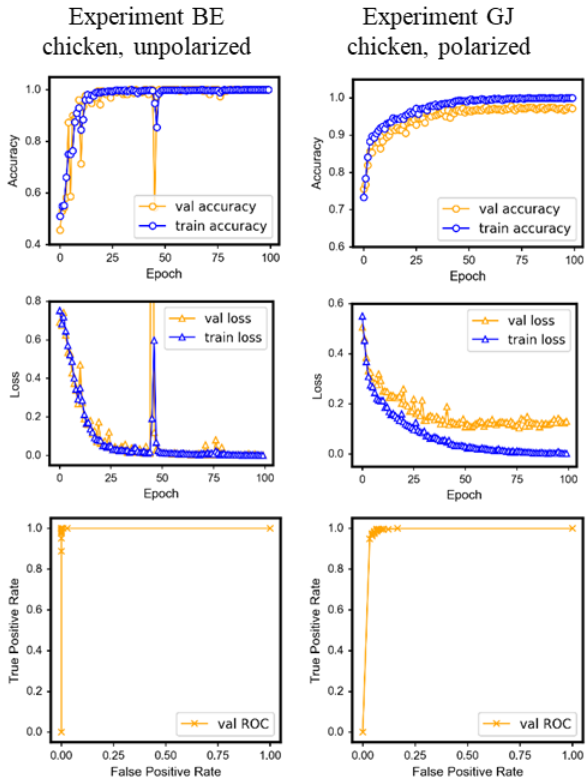
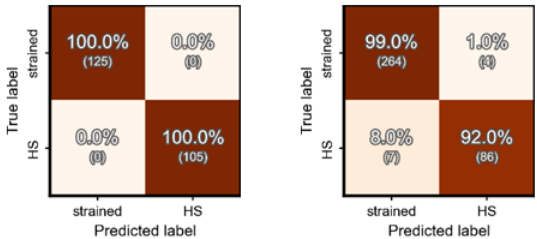
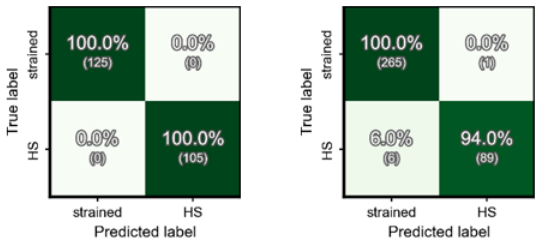

Figure 6. Results from experiments for the classification of healthy and HS (unhealthy) chicken RBCs. Each column (top to bottom) shows (1) accuracy and (2) loss achieved on training and validation sets; (3) ROC curve and corresponding (using best threshold) confusion matrix for the (4) validation set; and (5) test set.

Individual 'accuracy vs epoch curves' revealed that accuracy improved faster for unpolarized images when compared to polarized images. A training accuracy of 99% was achieved, as early as, at the end of epoch 6 for AB, epoch 13 for CD, epoch 49 for EG, and epoch 41 for HI. Thus, learning was more difficult on polarized images. Additionally, a training duration of 100 epochs seemed to be sufficient for achieving the highest possible accuracy (indicated by the plateau in the curve), except in the case of EG, where just the training (not validation) accuracy looked like it could improve further if the training wasn't stopped. Neither the accuracy nor the loss curves showed any quintessential signs of over- or under-fitting. The imbalance and the diversity in the sizes of the classes did not affect the learning in any noticeable way as observed here, in turn, demonstrating the applicability of this approach in a testing environment where it will not be feasible to collect much data or collect data in a balanced way.

The ROC curves helped determine a binary decision threshold (probability above which the trained model decides that a particular image sample belongs to a particular class) that yielded the highest possible true positive rate at the lowest possible false positive rate. In simpler terms, the best threshold is one for which a point on the ROC curve is closest to the point (0, 1). There was some variation in the appearance of the ROC curves, with the curvature increasing as training became more difficult, but most data points were still nicely bunched up towards the upper left corner and edges. The confusion matrices corresponding to this 'best threshold', for both the validation and test datasets revealed nothing out of the ordinary. Predictions were near perfect on the unpolarized images and at least 97% accurate on the polarized images, concurrent with the accuracies calculated by the model at the end of the training. A test accuracy of 100, 98.7, 97.7, and 98.1% was obtained for the experiments AB, CD, EG, and HI respectively.

**Classification of healthy and heat-stressed chicken RBCs:**

Figure 6 shows the results from experiments BE and GJ performed on chicken RBCs, that aimed towards the classification of healthy cells and cells experiencing HS (unhealthy) submerged in LC. Qualitatively, the results were similar to the results from the experiments in Figure 5. Accuracy and loss plots had similar shapes and approached similar values by the time the training ended. ROC plots were similar too. A validation accuracy of 100% was obtained for experiments using unpolarized images and greater than 92% for the experiments using polarized images. While the loss calculated for the training and validation sets were close to each other and reached zero for the unpolarized experiments, the loss for the validation set did not drop in sync with the training set for the polarized experiments and did not reach zero.

Here too, polarized images posed a greater challenge to the CNN algorithm. As described before, the color gradients around a cell, present in polarized images and absent in unpolarized imaged, are representative of the LC director being disrupted by the presence, the shape, and the physical properties of the healthy/HS cell, representative of cell's physical properties, its biochemical signatures (like the production of HSP70), and health. Because the classification labels were derived solely from the fact whether the cell was heat-stressed or not, the CNN's performance denoted the level at which it could successfully map the complex information present in the images (especially the polarized ones) to the effects of HS. In this light, the accuracy of 92/94 % for validation/test sets and higher was remarkable, given the myriad of humanly-indistinguishable information present in the images: these complexities presumably being greater than the effects of glutaraldehyde alone.

Similar to the glutaraldehyde crosslinked system, individual 'accuracy vs epoch curves' for healthy/HS cells revealed that accuracy improved faster for unpolarized images when compared to polarized images indicating that learning was more difficult on polarized images. A training duration of 100 epochs seemed to be sufficient for achieving the highest possible accuracy (indicated by the plateau in the curves). Neither the accuracy nor the loss curves, especially in conjunction with the ROC curves and the confusion matrices, showed any quintessential signs of over- or under-fitting. The dataset was still imbalanced and small for the healthy/HS pair but did not affect the learning in any noticeable way as observed here, in turn, demonstrating the applicability of this approach in a testing environment where it will not be feasible to collect much data or collect data in a balanced way.

The visual differences between healthy and crosslinked cells were more prominent than the differences between healthy and HS cells (Figure 4). We characterized the shapes of the healthy and heat-stressed chicken RBCs by quantifying the cell major ($r_x$) and minor axes ($r_y$). A higher aspect ratio ($r_x/r_y$) indicates a more significant strain. Chicken RBC, prior to straining, had an average aspect ratio of 1.5; upon applying mechanical strain, the aspect ratio increased to an average of 2.12 ± 0.64 for healthy cells and 1.85 ± 0.32 for heat-stressed cells. By comparing the variations in aspect ratio values of healthy and heat-stressed chicken RBCs, we observed that the healthy cells have a slightly higher strain than the heat-stressed cells however the datasets are not statistically different. Critically, the ML algorithm is able to identify the HS chicken cells with excellent accuracy (100% for unpolarized) even though the data sets are not statistically different.

## CONCLUSIONS

A simple, lightweight, convolutional neural network ML model, 3 convolution layers deep, was found to be capable of distinguishing between minute differences in the shapes or the aspect ratios of red blood cells (RBCs). The model was trained on microscope images of cells immersed in an isotonic solution of a nematic liquid crystal (DSCG). We demonstrate that LCs would 'sense' small differences in the mechanical properties of cells (or microbes), as evident from the changed shape of the cell or the liquid crystalline color pattern around the cells (in the polarized images) and could be used to detect the presence/onset of diseases, or microbes in air or water in tandem with trained ML models. An underlying and important hypothesis – that biochemical changes in an organism could affect mechanical changes in its cells – makes for a rich and interesting endeavor for future researchers; similar to how the expression of HSP70 protein was shown to make the heat-stressed RBCs stiffer in this study.

The 250K parameter model is easily trainable on GPU equipped personal computers using just a few hundred study-specific micrographs. The trained model is lightweight enough to fit in a sensor computer and fast enough to virtually instantly perform the classification on every new 'photograph'. In the context of the current rise of foundation vision models, whose fine-tuning and training are prohibitively expensive and possible only at very large research laboratories, our approach is geared towards empowering labs and individuals all round the world to construct and train their own ML models on general or niche tasks related to imaging and/or sensing.

In experiments, it was found that crosslinking the chicken and human RBCs using glutaraldehyde in order to simulate a diseased cell was an adequate strategy for planning, building, training, and evaluating valid ML models ahead of collecting of actual training data. In our case, no model-tweaking was found to be necessary while going from the simulated to the real heat-stressed cells. Because biological data could often be available in less quantities and later in the study, we believe that our simulation example might come in handy for researchers looking to work on their ML models while waiting, or those looking to simply augment their data in an appropriate manner.

## AUTHOR INFORMATION

**Corresponding Author**

* Karthik Nayani – Ralph E. Martin Department of Chemical Engineering, University of Arkansas, 72701, Fayetteville AR. Email: knayani@uark.edu

**Funding Sources**

KN acknowledges funding from USDA NIFA Award 2022-67021-36644.

**Notes**

The authors declare no competing financial interest.